\documentclass[runningheads]{llncs}
\usepackage{graphicx}

\usepackage{tikz}
\usepackage{comment}
\usepackage{amsmath,amssymb} 
\usepackage{color}
\usepackage{hyperref}
\usepackage{multirow}

\usepackage[accsupp]{axessibility}  

\begin{document}
\pagestyle{headings}
\mainmatter
\def\ECCVSubNumber{8}  

\title{MIPI 2022 Challenge on Quad-Bayer Re-mosaic: Dataset and Report} 

\titlerunning{MIPI 2022 Challenge on Quad-Bayer Re-mosaic}
%
\author{
Qingyu Yang \and 
Guang Yang \and
Jun Jiang \and
Chongyi Li \and
Ruicheng Feng \and
Shangchen Zhou \and
Wenxiu Sun \and
Qingpeng Zhu \and
Chen Change Loy \and
Jinwei Gu \and
Zhen Wang \and 
Daoyu Li \and  
Yuzhe Zhang \and 
Lintao Peng \and 
Xuyang Chang \and 
Yinuo Zhang \and
Yaqi Wu \and 
Xun Wu \and
Zhihao Fan \and
Chengjie Xia \and
Feng Zhang \and
Haijin Zeng \and
Kai Feng \and 
Yongqiang Zhao \and 
Hiep Quang Luong \and 
Jan Aelterman \and 
Anh Minh Truong \and
Wilfried Philips \and
Xiaohong Liu \and
Jun Jia \and 
Hanchi Sun \and 
Guangtao Zhai \and
Longan Xiao \and 
Qihang Xu \and
Ting Jiang \and
Qi Wu \and 
Chengzhi Jiang \and 
Mingyan Han \and 
Xinpeng Li \and 
Wenjie Lin \and 
Youwei Li \and 
Haoqiang Fan \and 
Shuaicheng Liu \and
Rongyuan Wu \and 
Lingchen Sun \and 
Qiaosi Yi  
\institute{~}
\vspace{-1cm}
}
\authorrunning{Q. Yang et al.}
%

\maketitle
\let\thefootnote\relax\footnotetext{\tiny Qingyu Yang$^{1}$ (\email{yangqingyu@sensebrain.site}), Jun Jiang$^{1}$ (\email{jiangjun@sensebrain.site}),  Chongyi Li$^{4}$, Shangchen Zhou$^{4}$, Ruicheng Feng$^{4}$, Wenxiu Sun$^{2,3}$, Qingpeng Zhu$^{2}$, Chen Change Loy$^{4}$, Jinwei Gu$^{1,3}$,   are the MIPI 2022 challenge organizers ($^{1}$SenseBrain, $^{2}$SenseTime Research and Tetras.AI, $^{3}$Shanghai AI Laboratory, $^{4}$Nanyang Technological University). The other authors participated in the challenge. Please refer to Appendix~\ref{appendix:teams} for details.\\ 
\\
MIPI 2022 challenge website: \url{http://mipi-challenge.org/}
}

\begin{abstract}
Developing and integrating advanced image sensors with novel algorithms in camera systems are prevalent with the increasing demand for computational photography and imaging on mobile platforms. However, the lack of high-quality data for research and the rare opportunity for in-depth exchange of views from industry and academia constrain the development of mobile intelligent photography and imaging (MIPI). To bridge the gap, we introduce the first MIPI challenge, including five tracks focusing on novel image sensors and imaging algorithms. In this paper, Quad Joint Remosaic and Denoise, one of the five tracks, working on the interpolation of Quad CFA to Bayer at full resolution, is introduced. The participants were provided a new dataset, including  70 (training) and 15 (validation) scenes of high-quality Quad and Bayer pairs. In addition, for each scene, Quad of different noise levels was provided at 0dB, 24dB, and 42dB. All the data were captured using a Quad sensor in both outdoor and indoor conditions. The final results are evaluated using objective metrics, including PSNR, SSIM~\cite{ssim}, LPIPS~\cite{lpips}, and KLD. A detailed description of all models developed in this challenge is provided in this paper. More details of this challenge and the link to the dataset can be found at \href{https://github.com/mipi-challenge/MIPI2022}{https://github.com/mipi-challenge/MIPI2022}

\keywords{Quad, Remosaic, Bayer, Denoise, MIPI challenge}
\end{abstract}

\section{Introduction}

\begin{figure}[!ht]
\centering
\includegraphics[width=0.5\textwidth]{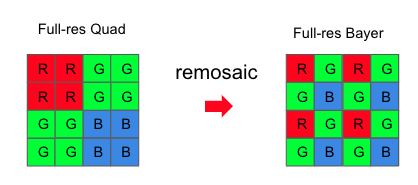}
\caption{The Quad remosaic task.}
\label{fig:quad}
\setlength{\belowcaptionskip}{0pt plus 3pt minus 2pt}
\end{figure}

Quad is a popular CFA pattern (Fig.~\ref{fig:quad}) widely used in smartphone cameras. Its binning mode is used under low light to enhance image quality by averaging four pixels within a $2\times2$ neighborhood. While signal-to-noise ratio (SNR) is improved in the binning mode, the spatial resolution is reduced as a tradeoff. To allow the output Bayer to have the same spatial resolution as the input Quad under normal lighting conditions, we need an interpolation procedure to convert Quad to a Bayer pattern. The interpolation process is usually referred to as remosaic. A good remosaic algorithm should be able to get the Bayer output from Quad with least artifacts, such as moire pattern, false color, and so forth.

The remosaic problem becomes more challenging when the input Quad becomes noisy. A joint remosaic and denoise task is thus in demand for real-world applications.

\begin{figure}[!ht]
\centering
\includegraphics[width=\textwidth]{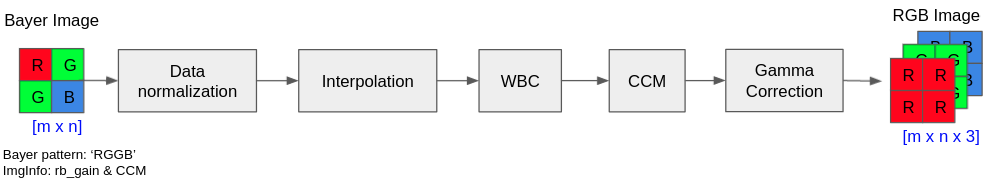}
\caption{An ISP to visuailize the output Bayer and to calculate the loss function.}
\label{fig:simple_isp}
\setlength{\belowcaptionskip}{0pt plus 3pt minus 2pt}
\end{figure}

In this challenge, we intend to remosaic the Quad input to obtain a Bayer at the same spatial resolution. The solution is not necessarily deep-learning. However, to facilitate the deep learning training, we provide a dataset of high-quality Quad and Bayer pairs, including 100 scenes (70 scenes for training, 15 for validation, and 15 for testing). We provide a Data Loader to read these files and show a simple ISP in Fig.~\ref{fig:simple_isp} to visualize the RGB output from the Bayer and to calculate loss functions. The participants are also allowed to use other public-domain datasets. The algorithm performance is evaluated and ranked using objective metrics: Peak Signal-to-Noise Ratio (PSNR), Structural Similarity Index (SSIM)~\cite{ssim}, Learned Perceptual Image Patch Similarity (LPIPS)~\cite{lpips}, and KL-divergence (KLD). The objective metrics of a baseline method are available as well to provide a benchmark. 

This challenge is a part of the Mobile Intelligent Photography and Imaging (MIPI) 2022 workshop and challenges emphasizing the integration of novel image sensors and imaging algorithms, which is held in conjunction with ECCV 2022. It consists of five competition tracks:
\begin{enumerate}
  \item RGB+ToF Depth Completion uses sparse and noisy ToF depth measurements with RGB images to obtain a complete depth map.
  \item Quad-Bayer Re-mosaic converts Quad-Bayer RAW data into Bayer format so that it can be processed by standard ISPs.
  \item RGBW Sensor Re-mosaic converts RGBW RAW data into Bayer format so that it can be processed by standard ISPs.
  \item RGBW Sensor Fusion fuses Bayer data and a monochrome channel data into Bayer format to increase SNR and spatial resolution.
  \item Under-display Camera Image Restoration improves the visual quality of the image captured by a new imaging system equipped with an under-display camera.
\end{enumerate}

\section{Challenge}
To develop a high-quality Quad Remosaic solution, we provide the following resources for participants:
\begin{itemize}
    \item A high-quality Quad and Bayer dataset: As far as we know, this is the first and only dataset consisting of aligned Quad and Bayer pairs, relieving the pain of data collection to develop learning-based remosaic algorithms;
    \item A data processing code with Data Loader to help participants get familiar with the provided dataset;
    \item A simple ISP including basic ISP blocks to visualize the algorithm output and to calculate the loss function on RGB results;
    \item A set of objective image quality metrics to measure the performance of a developed solution.
\end{itemize}

\subsection{Problem Definition}
Quad remosaic aims to interpolate the input Quad CFA pattern to obtain a Bayer of the same resolution. The remosaic task is needed mainly because current camera ISPs usually cannot process CFAs other than the Bayer pattern. In addition, the remosaic task becomes more challenging when the noise level gets higher, thus requiring more advanced algorithms to avoid image quality artifacts. In addition to the image quality requirement, Quad sensors are widely used in smartphones with a limited computational budget and battery life, thus requiring remosaic algorithm to be lightweight at the same time. While we do not rank solutions based on running time or memory footprint, the computational cost is one of the most important criteria in real applications.

\subsection{Dataset: Tetras-Quad}
\begin{figure}[!ht]
\centering
\includegraphics[width=\textwidth]{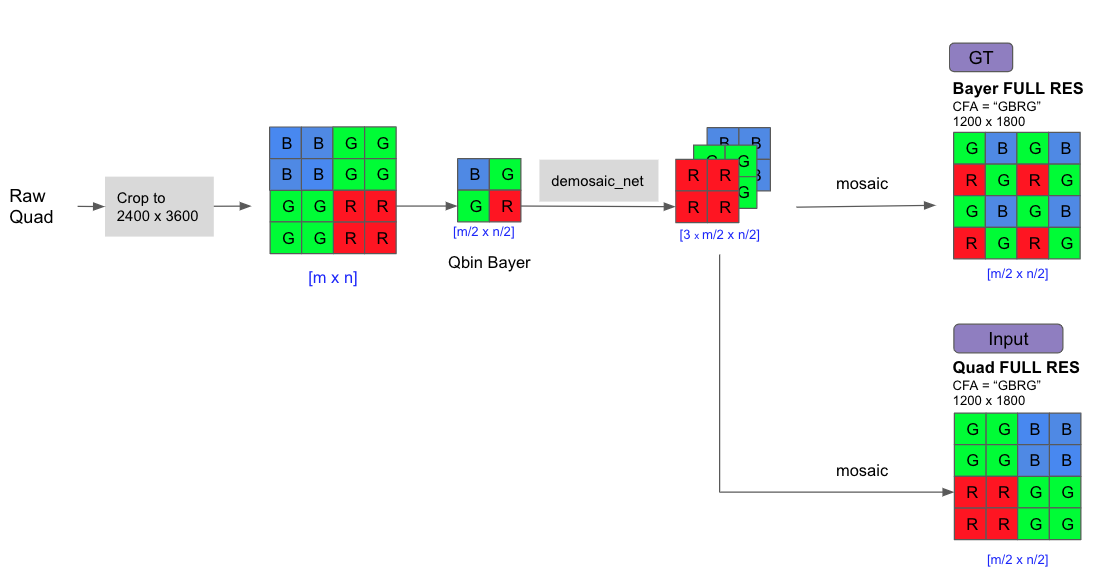}
\caption{Data generation of the Quad remosaic task. The Quad raw data is captured using a Quad sensor and cropped to be a size of $2400\times3600$. A Qbin Bayer is obtained by averaging a $2\times2$ block of the same color, and we demosaic the Bayer to get an RGB using the DemosaicNet~\cite{gharbi2016deep}. The RGB image is in turn mosaiced to get the input Quad and aligned with the ground truth Bayer.}
\label{fig:data_gen}
\setlength{\belowcaptionskip}{0pt plus 3pt minus 2pt}
\end{figure}

The training data contains 70 scenes of aligned Quad (input) and Bayer (ground-truth) pairs. For each scene, noise is synthesized on the 0dB Quad to provide the noisy Quad at 24dB and 42dB respectively. The synthesized noise consists of read noise and shot noise, and the noise models are measured on a Quad sensor. The data generation steps are shown in Fig.~\ref{fig:data_gen}. The testing data contains 15 Quad scenes at 0dB, 24dB, and 42dB, but the GT Bayer results are not available to participants. 

\subsection{Challenge Phases}
The challenge consisted of the following phases:
\begin{enumerate}
    \item Development: The registered participants get access to the data and baseline code, and are able to train the models and evaluate their running time locally.
    \item Validation: The participants can upload their models to the remote server to check the fidelity scores on the validation dataset, and to compare their results on the validation leaderboard.
    \item Testing: The participants submit their final results, code, models, and fact sheets.
\end{enumerate}

\subsection{Scoring System}
\subsubsection{Objective Evaluation}
The evaluation consists of (1) the comparison of the remosaic output (Bayer) with the reference ground truth Bayer, and (2) the comparison of RGB from the predicted and ground truth Bayer using a simple ISP (the code of the simple ISP is provided). We use
\begin{enumerate}
    \item Peak Signal-to-Noise Ratio (PSNR)
    \item Structural Similarity Index Measure (SSIM)~\cite{ssim}
    \item Kullback–Leibler Divergence (KLD)
    \item Learned Perceptual Image Patch Similarity (LPIPS)~\cite{lpips}
\end{enumerate}
to evaluate the remosaic performance. The PSNR, SSIM, and LPIPS will be applied to the RGB from the Bayer using the provided simple ISP code, while KLD is evaluated on the predicted Bayer directly.

A metric weighting PSNR, SSIM, KLD, and LPIPS is used to give the final ranking of each method, and we will report each metric separately as well. The code to calculate the metrics is provided. The weighted metric is shown below. The M4 score is between 0 and 100, and a higher score indicates a better overall image quality.

\begin{equation}
    \text{M4} = PSNR \cdot \text{SSIM}  \cdot 2^{1-\text{LPIPS}-\text{KLD}} .
\label{eq:M4}
\end{equation}
For each dataset we report the average results over all the processed images belonging to it.

\section{Challenge Results}

Six teams submitted their results in final phase, and their results have been verified using their submitted code as well. Table~\ref{tab:results} summarizes the results in the final test phase. \textbf{op-summer-po}, \textbf{JHC-SJTU}, and \textbf{IMEC-IPI \& NPU-MPI} are the top three teams ranked by M4 are presented in Eq.~\eqref{eq:M4}, and \textbf{op-summer-po} shows the best overall performance. The proposed methods are described in Section \ref{sec:methods} and the team members and affiliations are listed in Appendix \ref{appendix:teams}.

\begin{table}[!ht]  
    \centering
    
    \begin{tabular}{l | llll | l}
    \hline
        \textbf{Team name} & \textbf{PSNR} & \textbf{SSIM} & \textbf{LPIPS} & \textbf{KLD} &  \textbf{M4}\\ \hline  \hline
        \text{BITSpectral}                       & 37.2          & 0.96               & 0.11         & 0.03           &   66 \\ \hline
        \text{HITZST01}                         & 37.2           & 0.96               & 0.11         & 0.06           &   64.82 \\ \hline
        \text{IMEC-IPI \& NPU-MPI}      & 37.76         & 0.96               & 0.1          & 0.014          &  \textbf{67.95} \\ \hline
        \text{JHC-SJTU}                       & 37.64          & 0.96             & 0.1           & 0.007      &\textbf{67.99} \\ \hline
        \text{MegNR}                             & 36.08         & 0.95            & 0.095      & 0.023        &   64.1 \\ \hline
        \text{op-summer-po}               & 37.93          & 0.965           & 0.104       & 0.019        &  \textbf{68.03}  \\ \hline

    \end{tabular}
    \caption{MIPI 2022 Joint Quad Remosaic and Denoise challenge results and final rankings. PSNR, SSIM, LPIPS, and KLD are calculated between the submitted results from each team and the ground truth data. A weighted metric, M4, presented in Eq.~\eqref{eq:M4}, is used to rank the algorithm performance, and the team with the highest M4 is the winner. The M4 of the top 3 teams are highlighted.  
    \label{tab:results}}
\end{table}

\begin{figure}[!ht]
\setlength{\abovecaptionskip}{0.cm}
\setlength{\belowcaptionskip}{-0.cm}
\centering
\includegraphics[width=\textwidth]{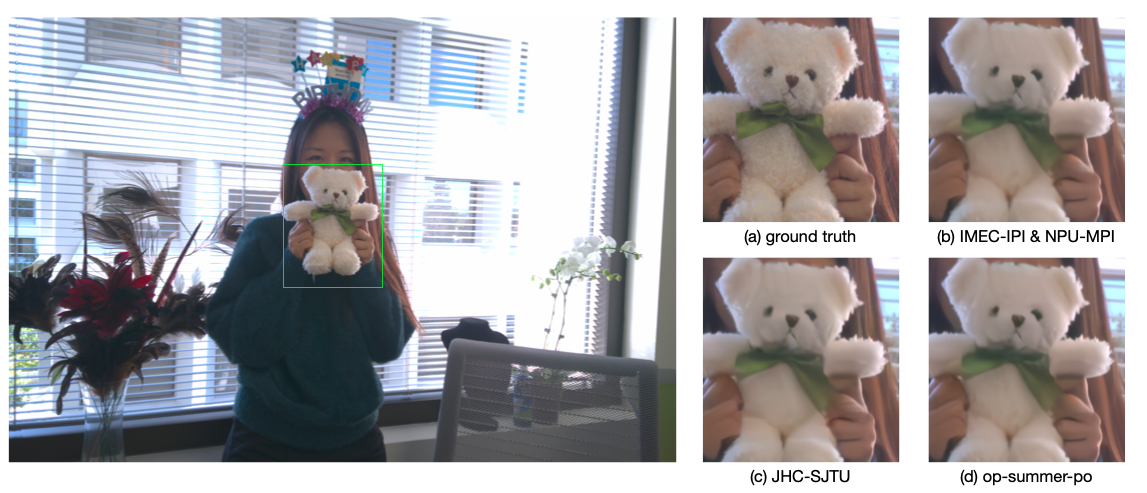}
\caption{Qualitative image quality (IQ) comparison. The results of one of the test scenes (42dB) are shown. While the top three remosaic methods achieve high objective IQ metrics in Table.~\ref{tab:results}, details and texture loss are noticeable on the Teddy bear. The RGB images are obtained by using the ISP in Fig.~\ref{fig:simple_isp}, and its code is provided to participants.}
\label{fig:IQ2}
\end{figure}

\begin{figure}[!ht]
\setlength{\abovecaptionskip}{0.cm}
\setlength{\belowcaptionskip}{-0.cm}
\centering
\includegraphics[width=\textwidth]{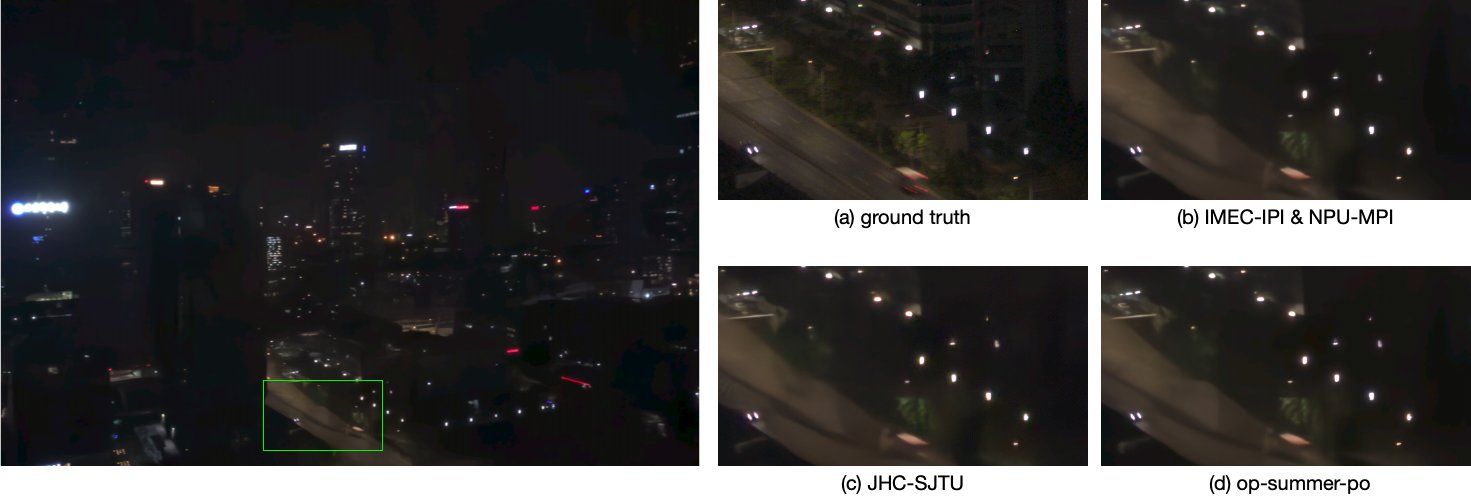}
\caption{Qualitative image quality (IQ) comparison. The results of one of the test scenes (42dB) are shown. Oversmoothing in the top three methods in Table.~\ref{tab:results} can be found when compared with the ground truth. The trees in the image can barely be identified in (b), (c), and (d). The RGB images are obtained by using the ISP in Fig.~\ref{fig:simple_isp}, and its code is provided to participants.}
\label{fig:IQ1}
\end{figure}

To learn more about the algorithm performance, we evaluated the qualitative image quality in addition to the objective IQ metrics in Figs.~\ref{fig:IQ2} and Fig.~\ref{fig:IQ1} respectively. 
While most teams in Table.~\ref{tab:results} have achieved high PSNR and SSIM, the detail and texture loss can be found on the Teddy bear in Fig.~\ref{fig:IQ2} and trees in Fig.~\ref{fig:IQ1}. When the input has a large amount of noise, oversmoothing tends to yield higher PSNR at the cost of detail loss perceptually.  

In addition to benchmarking the image quality of remosaic algorithms, computational efficiency is evaluated because of the wide adoption of Quad sensors in smartphones. We measured the runnnig time of the remosaic solutions of the top three teams (based on M4 by Eq.~\eqref{eq:M4}) in Table.~\ref{tab:runtime}. While running time is not employed in the challenge to rank remosaic algorithms, the computational cost is of critical importance when developing algorithms for smartphones. JHC-SJTU achieved the shortest running time among the top three solutions on a workstation GPU (NVIDIA Tesla V100-SXM2-32GB). With sensor resolution of mainstream smartphones reaching 64M or even higher, power-efficient remosaic algorithms are highly desirable.

\begin{table}[!ht]  
    \centering
    
    \begin{tabular}{l | l | l}
    \hline
        \textbf{Team name} & \textbf{1200$\times$1800 (measured)}  &  \textbf{64M} (estimated)\\ \hline  \hline
        \text{IMEC-IPI \& NPU-MPI}      & 6.1s                &  180s \\ \hline
        \text{JHC-SJTU}                        & \textbf{1s}     &  \textbf{29.6s} \\ \hline
        \text{op-summer-po}                & 4.4s               &  130s  \\ \hline

    \end{tabular}
    \caption{Running time of the top three solutions ranked by Eq.~\eqref{eq:M4} in the 2022 Joint Quad Remosaic and Denoise challenge. The running time of input of $1200\times1800$ was measured, while the running time of a 64M input Quad was based on estimation. The measurement was taken on an NVIDIA Tesla V100-SXM2-32GB GPU.
    \label{tab:runtime}}
\end{table}

\section{Challenge Methods}
\label{sec:methods}
In this section, we describe the solutions submitted by all teams paticipanting in the final stage of MIPI 2022 Joint Quad Remosaic and Denoise Challenge. 

\subsection{MegNR}
\begin{figure}[!ht]
\setlength{\abovecaptionskip}{0.cm}
\setlength{\belowcaptionskip}{-0.cm}
\centering
\includegraphics[width=0.9\textwidth]{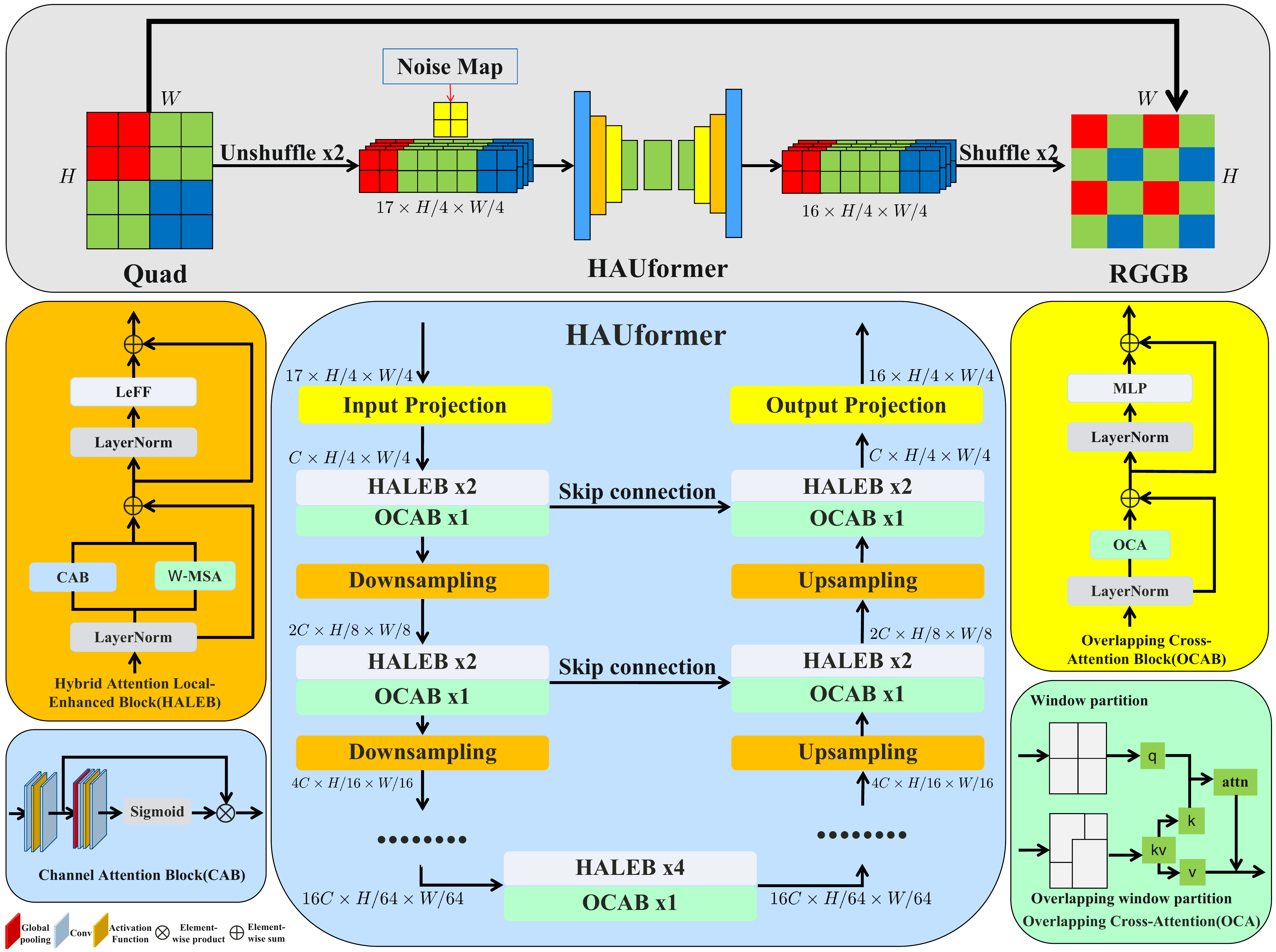}
\caption{The model architecture of MegNR.}
\label{fig:nn-MegNR}
\end{figure}
The overall pipeline is shown in Fig.~\ref{fig:nn-MegNR}. The Quad input is firstly split into independent channels by the pixel-unshuffle (PU)~\cite{sun2022hybrid}, which is served as a pre-processing module, and then the channels are fed into the Quad remosaic and reconstruction network, which is named as HAUformer similar to the Uformer~\cite{wang2022uformer}. The difference between them lies in the fact that they constructed two modules Hybrid Attention Local-Enhanced Block (HALEB) and Overlapping Cross-Attention Block (OCAB) to replace the LeWin Blocks~\cite{wang2022uformer} to capture more long-range dependencies information and useful local context. Finally, the pixel-shuffle (PS)~\cite{shi2016real} as a post-processing module is restored to the standard Bayer output.

The main contributions can be summarized as follows:
\begin{itemize}
    \item MegNR introduced channel attention and overlapping cross-attention module to Transformer to better aggregate input and cross-window information. \\
    \item A new loss function (Eq.~\eqref{eq:EqMegNR}) is proposed to constrain the network in multiple dimensions and significantly improve the network effect in Eq.~\eqref{eq:EqMegNR}. $\mathbf{M}$ denotes the loss function, $\mathbf{O_R}$ is the RGB output and $\mathbf{GT}$ is the ground truth. $\mathbf{PSNR}$, $\mathbf{SSIM}$, and $\mathbf{LPIPS}$ denote the visual perception metrics.\\
\end{itemize}
\begin{equation}\label{eq:EqMegNR}\small
\begin{aligned}
\mathbf{M} & = - \mathbf{P S N R}(\mathbf{O_R}, \mathbf{GT}) \cdot \mathbf{S S I M}(\mathbf{O_R}, \mathbf{GT}) \cdot 2^{1-\mathbf{L P I P S}(\mathbf{O_R}, \mathbf{GT})} .\\
\end{aligned}
\end{equation}

\subsection{IMEC-IPI \& NPU-MPI}
\begin{figure}[!ht]
\setlength{\abovecaptionskip}{0.cm}
\setlength{\belowcaptionskip}{-0.cm}
\centering
\includegraphics[width=0.99\textwidth]{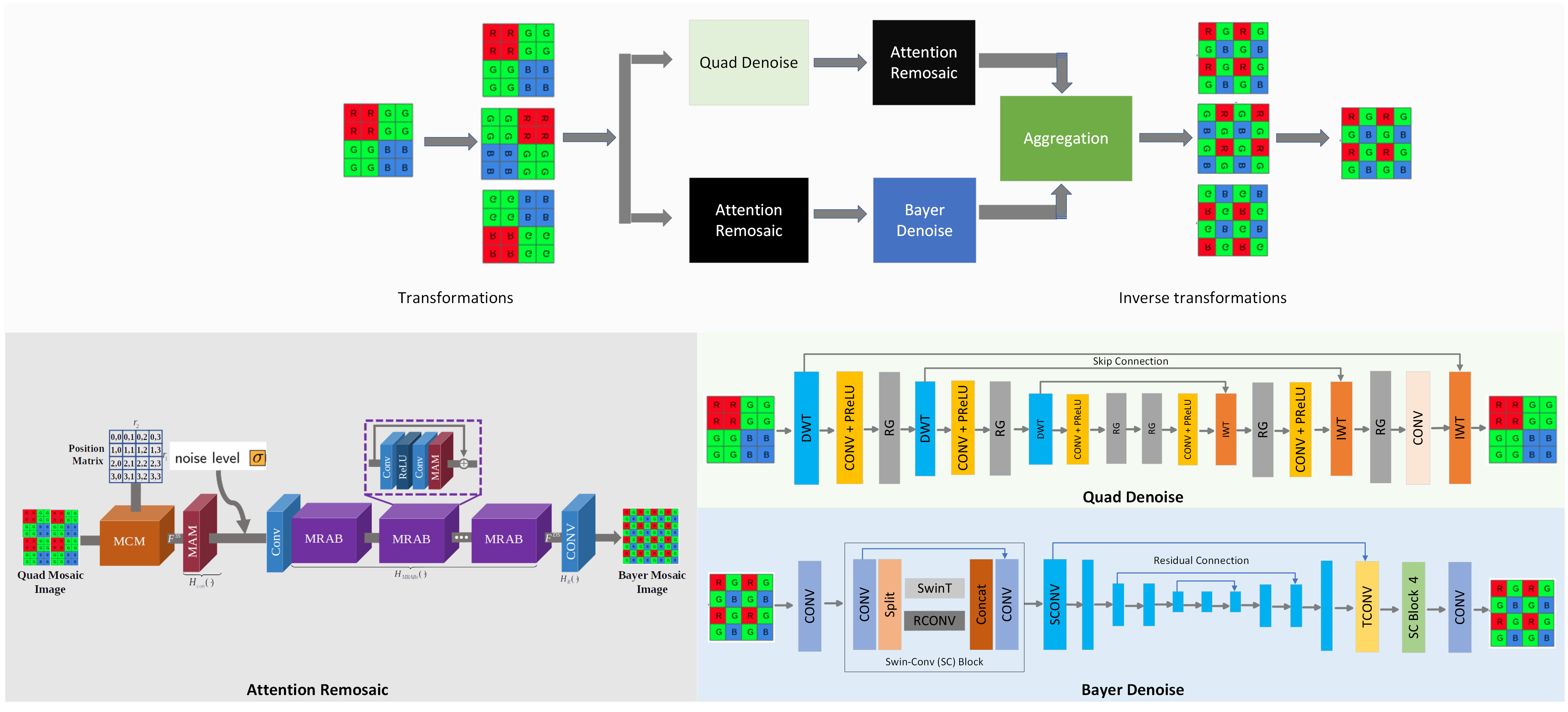}
\caption{The model architecture of IMEC-IPI \& NPU-MPI.}
\label{fig:nn-IMEC}
\end{figure}

\text{IMEC-IPI \& NPU-MPI} proposed a joint denoise and remosaic model, consisting of two transformation invariance guided parallel pipelines with independent Swin-Conv-UNet, a wavelet based denoising module, and mosaic attention based remosaicking module. The overall architecture is shown in Fig.~\ref{fig:nn-IMEC}.

Firstly, to avoid aliasing and artifacts in remosaicing, they utilize joint spatial-channel correlation in a raw Quad image by using a mosaic convolution module (MCM) and a mosaic attention module (MAM).

Secondly, for hard cases that the proposed naive mode trained using existing training datasets does not give satisfactory results, they proposed to identify difficult patches and create a better re-training set, leading to the improved performance on the hard cases.

Thirdly, for the denoising module, they employed the SCUNet~\cite{scunet} and DWT~\cite{Abdelhamed_2020_CVPR_Workshops} to address the realistic noise in this task. The SCUNet is used to denoise the Quad image, and they only trained it on the provided images without applying any pre-trained weights. The DWT is used to denoise the remosaiced Bayer image.

Finally, to combine denoising and remosaicking, they proposed a new parallel strategy and found the best way to reconstruct Bayer images from a noisy Quad mosaic, as shown in Fig.~\ref{fig:nn-IMEC}. In addition, to further improve the model robustness and generalization, they consider transformation invariance: a rigid transformation of the input should lead to an output with the same transformation. Here, they enforced this invariance through the network design instead of increasing the training dataset by direct rotating the training images.

\subsection{HITZST01}
\begin{figure}[!ht]
\setlength{\abovecaptionskip}{0.cm}
\setlength{\belowcaptionskip}{-0.cm}
\centering
\includegraphics[width=0.8\textwidth]{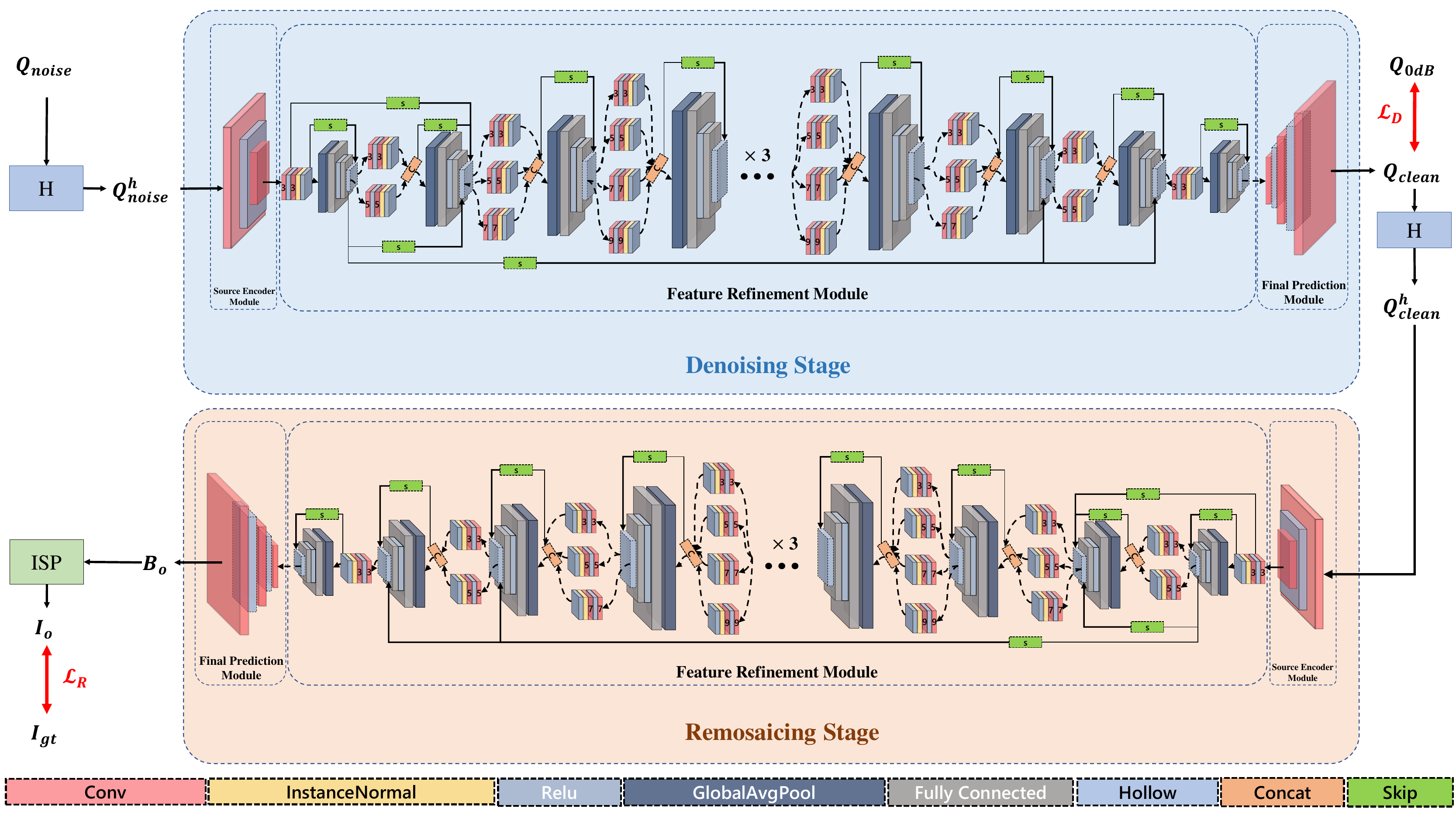}
\caption{The model architecture of HITZST.}
\label{fig:nn-HITZST}
\end{figure}
HITZST proposed a two-stage paradigm that formulating this task into denoise and remosaic. By introducing two supervisory signals: clean QBC $Q_{clean}$ for denoise stage and RGB images $I_{gt}$ generated from clean Bayer $B_{gt}$ for the remosaic stage, their entire framework is trained in a two-stage supervised manner in Fig.~\ref{fig:nn-HITZST},

Their proposed network generally consists of two stages: a denoise stage and a remosaic stage. For each stage, they used the same backbone~(JQNet). Specifically, their network framework consists of three sub-modules: the source encoder module, the feature refinement module, and the final prediction module.

The \textbf{Source Encoder Module} consists of two convolution layers. Its role is to reconstruct image texture information from mosaic image.
The \textbf{Feature Refinement Module} takes $I_{raw}$ as input to fully select, refine and enhance useful information in mosaic domain. It is formed by sequential connection of MCCA~\cite{PyNET}~\cite{zhang2018rcan} blocks of different scales. Moreover, \textbf{Final Prediction Module} takes $I_{refine}$ as the input and performs demosaicking to reconstruct a full-channel hyperspectral image.

\subsection{BITSpectral}
\begin{figure}[!ht]
\setlength{\abovecaptionskip}{0.cm}
\setlength{\belowcaptionskip}{-0.cm}
\centering
\includegraphics[width=0.8\textwidth]{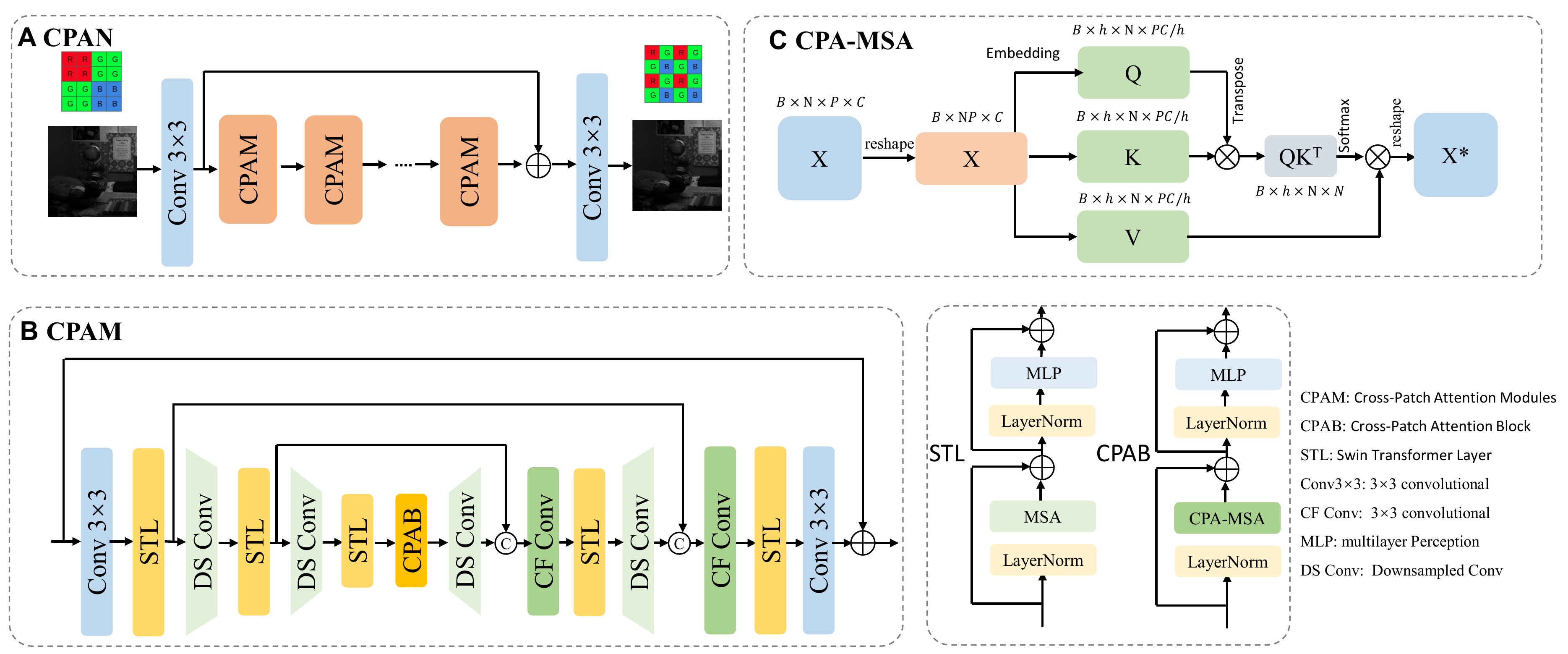}
\caption{The model architecture of BITSpectral.}
\label{fig:nn-BITSpectral}
\end{figure}
BITSpectral proposed a transformer-based network called CPAN in Fig.~\ref{fig:nn-BITSpectral}. The proposed Cross-Patch Attention Network (CPAN) follows the residual learning structure. The input convolutional layer extracts the embeddings of input images. The embeddings are next delivered into the Cross-Patch Attention Modules (CPAM). The reported network includes 5 CPAMs in total. Unlike conventional transformer modules, CPAM directly exploits the global attention between feature patches. CPAM is a U-shape sub-network and can reduce computational complexity. They utilized the Swin Transformer Layer (STL) to extract the attention within feature patches in each stage. The Cross-Patch Attention Block (CPAB) is used to directly obtain the global attention between patches for the innermost stage.

CPAB differs from STL in that it enhances the global field of perception by cross-patch attention. Current transformers in vision, such as Swin Transformer and ViT, perform self-attention within patches. They do not sufficiently consider the connection between patches, and thus the global feature extraction capability is limited.
CPAB consists of a Cross-Patch Attention Multi head Self-attention(CPA-MSA) and MLP layer. CPA-MSA is an improvement on traditional MSA, focusing on the relationship between different patches.

In the implementation, they divide the dataset into the training set (65 images) and the validation set (5 images). The images in the training set are cropped into patches of 128$\times$128 pixels. To improve efficiency, images in the testing set and validation set are not cropped.
They utilized the Adam optimizer with $\beta_1 = 0.9,\beta_2 = 0.999,e = 10^{-6}$ for training.
In order to accelerate the convergence rate, they applied the cosine annealing learning rate strategy, whose starting learning rate is set to $4 \times 10^{-4}$.
On a single 3090 GPU, the training procedure took around 20 hours to complete 100 epochs.
The CPAM-Net needs 0.08 seconds to reconstruct an image of $1200\times1800$ pixels in both validation and testing datasets.

\subsection{JHC-SJTU}
\begin{figure}[!ht]
\setlength{\abovecaptionskip}{0.cm}
\setlength{\belowcaptionskip}{-0.cm}
\centering
\includegraphics[width=0.8\textwidth]{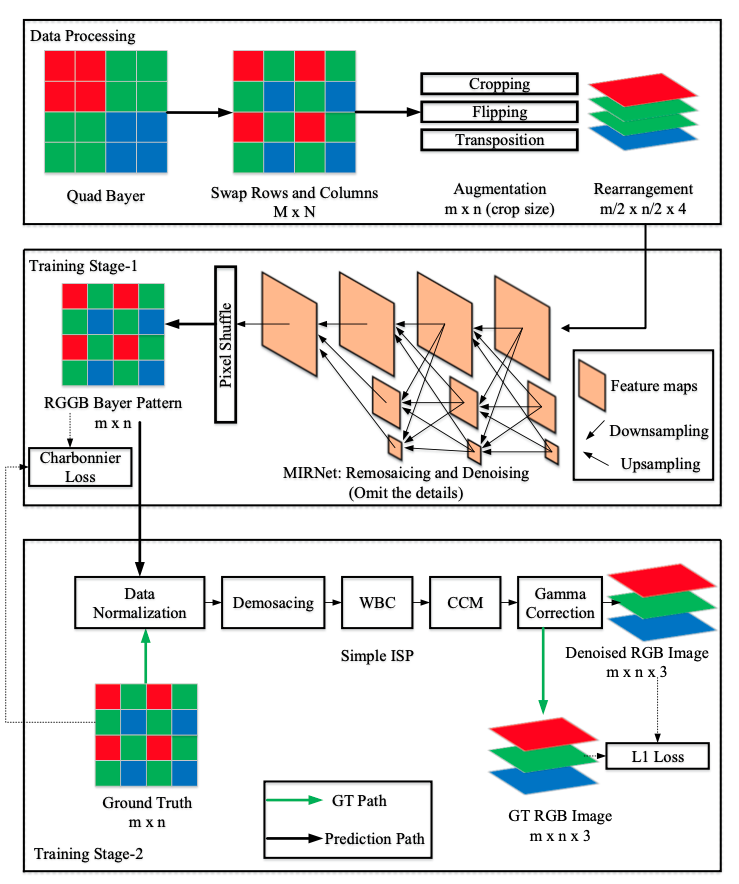}
\caption{The model architecture of JHC-SJTU.}
\label{fig:nn-JHC-SJTU}
\end{figure}
JHC-SJTU proposed a joint Quad remosaicing and denoising pipeline. The pipeline consists of two stages as shown in Fig.~\ref{fig:nn-JHC-SJTU}. In the first stage, a multi-resolution feature extraction and fusion network is used to jointly remosaic the Quad to standard Bayer (RGGB) and denoise it. The ground truth raw data in RGGB pattern is used for supervision. In the second stage, the simple ISP provided by the organizers is concatenated to the output of the network. They fine-tuned the network in the first stage using the RGB images generated by the ISP without updating its parameters, i.e. the demosaicing network~\cite{gharbi2016deep}. The jointly remosaicing and denosing network is MIRNet~\cite{Zamir2020MIRNet}. The main contributions of the proposed solution include:
\begin{itemize}
    \item They proposed a prepocessing method: for Quad raw data, they first swap the second column and the third column in each 4$\times$4 Quad unit, and then swap the second row and the third row of each unit. After that, the Quad raw data is converted to an RGGB Bayer. Then, they decomposed the RGGB into four channel maps that are the R, G1, G2, and B, respectively. With these prepocessing steps, they could extract the channel information collected from different sensors.
\\
    \item 
After the training in the first stage, they concatenated the trained ISP to the joint remosaicing and denoising network. The output of their network and the ground truth RGGB raw data were processed to generate RGB images. Then, the generated ground-truth images were used as a color-supervision to fine-tune their network. In this stage, they freezed the parameters of the ISP. Compared to raw data, RGB images contain more information, which can further improve the quality of the generated raw data. Their experiments demonstrated the improvement in PSNR is about 0.2 dB.
\\
    \item They also used data augmentation to expand the training samples. In training, cropping, flipping, and transposition were randomly applied to the raw data. In addition, they designed a cropping-based method to unify the Bayer pattern after these augmentations. The experimental results showed these augmentation methods could increase the PSNR by about 0.4 dB.

\end{itemize}

\subsection{op-summer-po}
\begin{figure}[!ht]
\setlength{\abovecaptionskip}{0.cm}
\setlength{\belowcaptionskip}{-0.cm}
\centering
\includegraphics[width=0.8\textwidth]{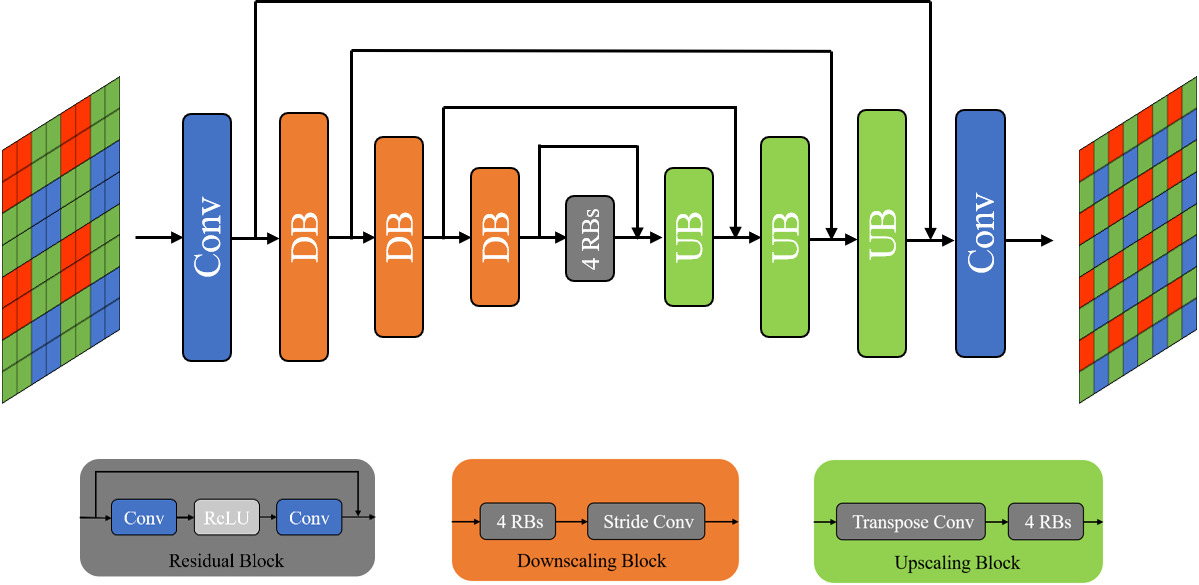}
\caption{The model architecture of op-summer-po.}
\label{fig:nn-op-summer-po}
\end{figure}
Op-summer-po proposed to do the Quad remosaic and denoise using DRUNet\cite{zhang2021plug}. The overall model architecture is shown in Fig.~\ref{fig:nn-op-summer-po}. The DRUNet has four scales with 64, 128, 256, and 512 channels respectively. Each scale has an identity skip connection between the stride convolution (stride = 2) and the transpose convolution. This connection concatenates encoder and decoder features. Each encoder or decoder includes four residual blocks. 

To avoid detail loss during denoising, LPIPS\cite{lpips} is used both on the raw and RGB images in the training. No pre-trained models or external datasets were used for the task. During training, the resolution of HR patches is 128 × 128. They implemented the experiment on 1*V100 with PyTorch and the batch size is 16 per GPU. The optimizer is Adam. The learning rate was initially set to 0.0001, and it decayed by 0.5.

\section{Conclusions}
In this paper, we summarized the Joint Quad Remosaic and Denoise challenge in the first Mobile Intelligent Photography and Imaging workshop (MIPI 2022) held in conjunction with ECCV 2022. The participants were provided with a high-quality training/testing dataset for Quad remosaic and denoise, which is now available for researchers to download for future research. We are excited to see that so many submissions within such a short period, and we look forward for more research in this area.

\section{Acknowledgements}
We thank Shanghai Artificial Intelligence Laboratory, Sony, and Nanyang Technological University to sponsor this MIPI 2022 challenge. We thank all the organizers and participants for their great work. 

\appendix
\section{Teams and Affiliations}
\label{appendix:teams}
\tiny
\textbf{BITSpectral} \\
\textbf{Title}: Cross-Patch Attention Network for Quad Joint Remosaic and Denoise \\
\textbf{Members}:
Zhen Wang, (wzhstruggle@bit.edu.cn), Daoyu Li,  Yuzhe Zhang, Lintao Peng, Xuyang Chang, Yinuo Zhang \\
\textbf{Affiliations}:
Beijing Institute of Technology
\\
\\
\textbf{HITZST01} \\
\textbf{Title}: Multi-scale convolution network (JDRMCANet) with joint denoise and remosaic.\\
\textbf{Members}:
$^{1}$Yaqi Wu (titimasta@163.com), 
$^{2}$Xun Wu,
$^{3}$Zhihao Fan,
$^{4}$Chengjie Xia,
Feng Zhang,\\
\textbf{Affiliations}:
$^{1}$Harbin Institute of Technology, Harbin, 150001, China, $^{2}$Tsinghua University, Beijing, 100084, China,
$^{3}$University of Shanghai for Science and Technology, Shanghai, 200093, China, $^{4}$Zhejiang University, Hangzhou, 310027, China
\\
\\
\textbf{IMEC-IPI \& NPU-MPI}\\
\textbf{Title}: Hard Datasets and Invariance guided Parallel Swin-Conv-Attention Network for Quad Joint Remosaic and Denoise.\\
\textbf{Members}:
$^{1}$Haijin Zeng (Haijin.Zeng@imec.be), $^{2}$Kai Feng, $^{2}$Yongqiang Zhao, $^{1}$Hiep Quang Luong, $^{1}$Jan Aelterman, $^{1}$Anh Minh Truong, and $^{1}$Wilfried Philips.\\
\textbf{Affiliations}:
$^{1}$IMEC \& Ghent University
$^{2}$Northwestern Polytechnical University
\\
\\
\textbf{JHC-SJTU}\\
\textbf{Title}: Jointly Remosaicing and Denoising for Quad Bayer with Multi-Resolution Feature Extraction and Fusion.\\
\textbf{Members}:
$^{1}$Xiaohong Liu (xiaohongliu@sjtu.edu.cn), $^{1}$Jun Jia, $^{1}$Hanchi Sun, $^{1}$Guangtao Zhai, $^{2}$Anlong Xiao, $^{2}$Qihang Xu \\
\textbf{Affiliations}:
$^{1}$Shanghai Jiao Tong University, $^{2}$Transsion\\
\\
\\
\textbf{MegNR}\\
\textbf{Title}: HAUformer: Hybrid Attention-guided U-shaped Transformer for Quad Remosaic Image Restoration.\\
\textbf{Members}:
Ting Jiang (jiangting@megvii.com), Qi Wu, Chengzhi Jiang, Mingyan Han, Xinpeng Li, Wenjie Lin, Youwei Li, Haoqiang Fan and Shuaicheng Liu\\
\textbf{Affiliations}:
Megvii Technology\\
\\
\\
\textbf{op-summer-po}\\
\textbf{Title}: Two LPIPS Functions in Raw and RGB domains for Quad-Bayer Joint Remosaic and Denoise.\\
\textbf{Members}:
$^{1}$Rongyuan Wu (1104138645@qq.com), $^{1}$Lingchen Sun, $^{1,2}$Qiaosi Yi\\
\textbf{Affiliations}:
$^{1}$OPPO Research Institute, $^{2}$East China Normal University\\
\\
\\

%
%
\bibliographystyle{splncs04}
\bibliography{egbib}

\begin{thebibliography}{10}
\providecommand{\url}[1]{\texttt{#1}}
\providecommand{\urlprefix}{URL }
\providecommand{\doi}[1]{https://doi.org/#1}

\bibitem{Abdelhamed_2020_CVPR_Workshops}
Abdelhamed, A., Afifi, M., Timofte, R., Brown, M.S.: Ntire 2020 challenge on
  real image denoising: Dataset, methods and results. Proceedings of the
  IEEE/CVF Conference on Computer Vision and Pattern Recognition Workshops
  (CVPRW)  (2020)

\bibitem{gharbi2016deep}
Gharbi, M., Chaurasia, G., Paris, S., Durand, F.: Deep joint demosaicking and
  denoising. ACM Transactions on Graphics (ToG)  \textbf{35}(6),  1--12 (2016)

\bibitem{PyNET}
Kim, B.H., Song, J., Ye, J.C., Baek, J.: Pynet-ca: enhanced pynet with channel
  attention for end-to-end mobile image signal processing. European Conference
  on Computer Vision (ECCV) pp. 202--212 (2020)

\bibitem{shi2016real}
Shi, W., Caballero, J., Husz{\'a}r, F., Totz, J., Aitken, A.P., Bishop, R.,
  Rueckert, D., Wang, Z.: Real-time single image and video super-resolution
  using an efficient sub-pixel convolutional neural network. Proceedings of the
  IEEE/CVF Conference on Computer Vision and Pattern Recognition (CVPR) pp.
  1874--1883 (2016)

\bibitem{sun2022hybrid}
Sun, B., Zhang, Y., Jiang, S., Fu, Y.: Hybrid pixel-unshuffled network for
  lightweight image super-resolution. arXiv preprint arXiv:2203.08921  (2022)

\bibitem{wang2022uformer}
Wang, Z., Cun, X., Bao, J., Zhou, W., Liu, J., Li, H.: Uformer: A general
  u-shaped transformer for image restoration. Proceedings of the IEEE/CVF
  Conference on Computer Vision and Pattern Recognition (CVPR) pp. 17683--17693
  (2022)

\bibitem{ssim}
Wang, Z., Bovik, A.C., Sheikh, H.R., Simoncelli, E.P.: Image quality
  assessment: from error visibility to structural similarity. IEEE transactions
  on image processing  \textbf{13}(4),  600--612 (2004)

\bibitem{Zamir2020MIRNet}
Zamir, S.W., Arora, A., Khan, S., Hayat, M., Khan, F.S., Yang, M.H., Shao, L.:
  Learning enriched features for real image restoration and enhancement.
  European Conference on Computer Vision (ECCV)  (2020)

\bibitem{scunet}
Zhang, K., Li, Y., Liang, J., Cao, J., Zhang, Y., Tang, H., Timofte, R.,
  Van~Gool, L.: Practical blind denoising via swin-conv-unet and data
  synthesis. arXiv preprint  (2022)

\bibitem{zhang2021plug}
Zhang, K., Li, Y., Zuo, W., Zhang, L., Van~Gool, L., Timofte, R.: Plug-and-play
  image restoration with deep denoiser prior. IEEE Transactions on Pattern
  Analysis and Machine Intelligence  (2021)

\bibitem{lpips}
Zhang, R., Isola, P., Efros, A.A., Shechtman, E., Wang, O.: The unreasonable
  effectiveness of deep features as a perceptual metric. Proceedings of the
  IEEE/CVF Conference on Computer Vision and Pattern Recognition Workshops
  (CVPRW)  (2018)

\bibitem{zhang2018rcan}
Zhang, Y., Li, K., Li, K., Wang, L., Zhong, B., Fu, Y.: Image super-resolution
  using very deep residual channel attention networks. European Conference on
  Computer Vision (ECCV)  (2018)

\end{thebibliography}
\end{document}